\documentclass[showpacs,showkeys]{revtex4}%
\usepackage[colorlinks=true]{hyperref}
\AtBeginDocument{
	\hypersetup{
		urlcolor=cyan,
		citecolor=magenta,
		linkcolor=blue,
		anchorcolor=green,
	}%
}
\usepackage{amssymb}
\usepackage{amsmath}
\usepackage{amsfonts}
\usepackage{graphicx}
\usepackage{ulem} 
\usepackage{cancel}
\usepackage[usenames]{color}
\usepackage{epstopdf}
\usepackage{extarrows}
\providecommand{\rbr}[1]{\left( #1 \right)}%
\providecommand{\sqbr}[1]{\left[ #1 \right]} %

\def\ds{\displaystyle}

\begin{document}

\title[ ]{Entropy Maximization with Linear Constraints: The Uniqueness of the Shannon Entropy}
\author{$^{1}$Thomas Oikonomou}
\email{thomas.oikonomou@nu.edu.kz}
\author{$^{2}$G. Baris Bagci}
\affiliation{$^{1}$Department of Physics, School of Science and Technology, Nazarbayev University, Astana 010000, Kazakhstan}
\affiliation{$^{2}$Department of Materials Science and Nanotechnology Engineering, TOBB University of Economics and Technology, 06560 Ankara, Turkey}

\begin{abstract}
Within a framework of utmost generality, we show that the entropy maximization procedure with linear constraints uniquely leads to the Shannon-Boltzmann-Gibbs entropy. Therefore, the use of this procedure with linear constraints should not be extended to the generalized entropies introduced recently. 
In passing, it is remarked how the forceful use of the entropy maximization for the Tsallis and R\'enyi entropies implies either the Shannon limit of these entropies or self-referential contradictions.
Finally, we note that the utilization of the entropy maximization procedure with different averaging schemes is beyond the scope of this work.    
\end{abstract}

\eid{ }
\date{\today }
\startpage{1}
\endpage{1}
\maketitle

\section{Introduction}
%

Since Jaynes \cite{Jaynes1}, the entropy maximization has been a widely used tool in many different fields benefiting from Shannon entropy. Although the initial aim of Jaynes was to derive the equilibrium distribution associated with the Shannon entropy subject to the linear constraints, recent progress in the generalized entropies such as Tsallis \cite{Tsallis1988} or R\'enyi \cite{Renyi} entropies, to mention but a few, also benefited from the very same entropy maximization procedure with various applications \cite{Bagci1,Rotundo,Van1,Van2,Wong,reis,
Portesi,Bagci2,Campisi,Oik2007,high3,third,Rob1,esc3}.

However, we have recently shown that the entropy maximization with linear constraints does not yield a distribution which can be cast into the appropriate form so as to include the partition function when it is used for the generalized entropies \cite{Th}. In other words, the distributions are not of the form $p_i=f^{-1}(\beta \varepsilon_i)/\sum_k f^{-1}(\beta \varepsilon_k)$ ($\beta$ being the Lagrange multiplier of the internal energy constraint and $\varepsilon_i$ is the energy of the $i$th micro state) so that the the denominator (i.e. normalization term) could not be identified as the partition function. The sole possibility for such a distribution has been found to be the one associated with the Shannon entropy.

In this work, we do not interest ourselves with the explicit form of the distribution. Instead, in its all generality, we consider the entropy maximization with linear constraints as Jaynes previously did \cite{Jaynes1} and show that the only admissible entropy expression is the Shannon (or Boltzmann-Gibbs) entropy. Therefore, we point out that the entropy maximization construed {\`a} la Jaynes is suitable only for the Shannon entropy and thereby excludes the use of any generalized entropies.

\section{Maximization procedure revisited}\label{SecII}

The entropy functional with linear constraints reads
 
\begin{eqnarray}\label{eq02}
L(\{p\},\alpha,\beta,U)=S(\{p\})-\alpha\sqbr{\sum_{i=1}^{n} p_i-1} - \beta\sqbr{\sum_{i=1}^{n} p_i\varepsilon_i-U}\,,
\end{eqnarray}
where $S$ denotes the entropy measure and $U$ is the internal energy. As usual, $p_i$ is the probability of occurrence for the $i$th micro state and $(\alpha,\beta)$ are the respective Lagrange multipliers. Considering the maximization functional in Eq. (\ref{eq02}) and using the definition $\partial S(\{p\})/\partial p_i=:f(p_i)$, the maximization procedure yields the following $n+3$ equations \cite{Karabulut}
\begin{subequations}\label{eq03}
\begin{eqnarray}\label{eq03a}
f(p_i)&=&\alpha + \beta \varepsilon_i=x_{i}\,,\\
\label{eq03b}
1&=&\sum_{i=1}^{n} p_i\,,\\
\label{eq03c}
U&=&\sum_{i=1}^{n} p_i\varepsilon_i\,,\\
\label{eq03d}
\beta &=& \frac{\partial S}{\partial U}\,.
\end{eqnarray}
\end{subequations}

Taking the derivative of Eq. (\ref{eq03b}) with respect to $\beta$, we have
\begin{eqnarray}\label{alphaRelation}
0&=& \sum_{i=1}^{n}\frac{\partial p_i}{\partial \beta}
=\sum_{i=1}^{n}\frac{\partial f^{-1}(\alpha+\beta\varepsilon_i)}{\partial \beta}
=\sum_{i=1}^{n}\frac{\partial f^{-1}(\alpha+\beta\varepsilon_i)}{\partial (\alpha+\beta\varepsilon_i)}\frac{\partial(\alpha+\beta\varepsilon_i)}{\partial\beta}\,.
\end{eqnarray}
Introducing the normalized $P_i$ as
\begin{eqnarray}\label{BigP}
P_i= \rbr{\sum_{k=1}^{n}\frac{\partial f^{-1}(x_k)}{\partial  x_k}}^{-1}\frac{\partial f^{-1}(x_i)}{\partial  x_i}\,,
\end{eqnarray}
Eq. (\ref{alphaRelation}) yields
\begin{eqnarray}\label{delalpha}
\frac{\partial \alpha}{\partial \beta} &=& - \sum_{i=1}^{n} P_i\varepsilon_i=-\widetilde{U}\,.
\end{eqnarray}
The quantity $\widetilde{U}$ is related to $U$ as (combine Eqs. (\ref{eq03})- (\ref{delalpha}))
\begin{eqnarray}
\label{eqNew01}
\widetilde{U}&=& U - \sum_{i=1}^{n}p_i\frac{\partial f(p_i)}{\partial\beta}\,.
\end{eqnarray}
Similarly to Eq. (\ref{alphaRelation}), since $P_i$ satisfies the normalization condition, we have
\begin{eqnarray}\label{delbeta}
0=\sum_{i=1}^{n} \frac{\partial P_i}{\partial \beta}= \sum_{i=1}^{n} \frac{\partial P_i}{\partial x_i}\left(\varepsilon_i - \widetilde{U}\right)\quad\Rightarrow\quad
\widetilde{U}=\sum_{i=1}^{n}\frac{\frac{\partial P_i}{\partial x_i}}{\sum_{k=1}^{n}\frac{\partial P_k}{\partial x_k}}\varepsilon_i\,.
\end{eqnarray}
Comparing Eqs. (\ref{delalpha}) and (\ref{delbeta}) we read
\begin{eqnarray}\label{DefY_i}
\sum_{i=1}^{n}Y_i\varepsilon_i=0\,,\qquad 
Y_i:= P_i -  \frac{\frac{\partial P_i}{\partial x_i}}{\sum_{k=1}^{n}\frac{\partial P_k}{\partial x_k}}\,.
\end{eqnarray}
The validity of this equation presents us with two cases we inspect below:\\

\noindent(i.) The first possibility, assuming $Y_i\neq0$, is that the total sum can be equal to zero. Then, applying the $m$th derivative with respect to $\beta$ yields
\begin{eqnarray}\label{eqA3}
\sum_{i=1}^{n}\frac{\partial^m Y_i}{\partial \beta^m} \varepsilon_i=0
\end{eqnarray}
This is a $m\times n$ homogeneous system of the form $A_{ij}X_i=0$ ($i=1,\ldots,n$, $j=1,\ldots,m$) to be solved with $A_{ij}\equiv \frac{\partial^{j} Y_j}{\partial \beta^{j}}$ and $X_i\equiv\varepsilon_i$. Then, we know from linear algebra that the former system has either the zero solution, i.e.,  $X_i=0$, or  a set of infinite solutions with $A_{ij}=A_{i\ell}$. The zero solution is apparently not an option. Thus, we have infinite solutions yielding $\frac{\partial^{j}}{\partial \beta^{j}}Y_i = \frac{\partial^{\ell}}{\partial\beta^{\ell}}Y_i\;\Rightarrow\; Y_i=ce^{\beta}$. Summing over all $i$'s and using the normalization condition we see that the former relation is only possible when  $c=0\;\Rightarrow\;Y_i=0$, which is a contradiction to our initial assumption.\\

\noindent(ii.) The second and only possibility that is left is 
\begin{eqnarray}\label{eq10}
Y_i=0\,.
\end{eqnarray}
Then,  substituting the definition of $Y_i$ in Eq. (\ref{DefY_i}) into the former equality, we have
\begin{eqnarray}\label{important}
\frac{\partial}{\partial x_i} \ln(P_i) =
\sum_{k=1}^{n} \frac{\partial P_k}{\partial x_k}
\end{eqnarray}
Since the l.h.s. and r.h.s. have an open and a closed $i$ dependence (or equivalently, the former depends and the latter does not depend on $i$), respectively, the only option satisfying this relation is $\ln(P_i)\sim x_i$ so that the derivative eliminates the $i$-dependence. Thus, the only option is that the measure $P_i$ has to be the inverse logarithmic function, i.e., 
\begin{eqnarray}\label{ffun}
P_i= \exp\rbr{\ds\pm\frac{x_i}{k}}\,,
\end{eqnarray}
where $k$ is merely a constant. By virtue of Eq. (\ref{ffun}), we read in Eq. (\ref{BigP})
\begin{eqnarray}\label{ffun2}
\sum_{k=1}^{n}\frac{\partial f^{-1}(x_k)}{\partial x_k} = \exp\left(\mp\frac{x_i}{k}\right)\frac{\partial f^{-1}(x_i)}{\partial x_i}\,.
\end{eqnarray}
Then, a similar discussion to Eq. (\ref{important}) uniquely yields $P_i=f^{-1}(x_i)=p_i$, hence
\begin{eqnarray}\label{ffun3}
f^{-1}(x_i)= \exp\rbr{\ds\pm\frac{x_i}{k}} \qquad\Leftrightarrow\qquad
f(p_i)=\pm k\ln(p_i)\,.
\end{eqnarray}

To reiterate, the MaxEnt procedure with linear constraints leads to two distinct, at first glance,  probability distribution sets, $\{p_i\}$ and $\{P_i\}$, respectively. The former is used in the maximization procedure itself and the latter was deduced from the normalization condition of $p_i$. However, the normalization of $P_i$ in turn shows that these two distribution sets are actually one and the same, $P_i=p_i\;\Rightarrow\; U=\widetilde{U}$, exhibiting an exponential behavior with respect to the energy values $\varepsilon_i$.

\section{Determining the Entropy Uniquely}

We now show how considerations in the previous section uniquely leads to the Shannon-Boltzmann-Gibbs entropy. Integrating Eq. (\ref{eq03d}) with respect to $U$, we have
\begin{eqnarray}\label{entropy01}
S=\beta U - \int U\mathrm{d}\beta+C_1\,,
\end{eqnarray}
where $C_1$ is the integration constant and does not depend on $\beta$. Using the mean value constraint in Eq. (\ref{eq03c})  the former equation can be written as
\begin{eqnarray}\label{entropy02}
S=\sum_{i=1}^{n} p_i(\beta \varepsilon_i) -\int U\mathrm{d}\beta +C_1\,,
\end{eqnarray}
Taking into account Eqs. (\ref{eq03a}) and (\ref{ffun}) and then Eqs. (\ref{delalpha}) and (\ref{eq10}), Eq. (\ref{entropy02}) can be written as
\begin{eqnarray}\label{entropy03}
S=\pm k \sum_{i=1}^{n}p_i\ln(p_i)  + C\,.
\end{eqnarray}
This is the most general structure of the entropy $S$ satisfying the MaxEnt procedure with linear constraints. 
The term $C$ includes all additive constants. The sign in Eq. (\ref{entropy03}) depends on whether the entropy $S$ is to be maximized  or minimized (negative or positive sign, respectively).
For $k=1$ this is identified with the Shannon entropy and for $k=k_\text{\tiny{B}}$ with the Boltzmann-Gibbs entropy within the information theory and statistical thermodynamics, respectively.

\section{Conclusions}

Since the seminal work of Jaynes \cite{Jaynes1}, entropy maximization procedure has been utilized in the literature. However, in the recent decades, this procedure has been used for various entropy definitions such as Tsallis \cite{Tsallis1988} or R{\'e}nyi entropies \cite{Renyi}, although Jaynes originally used only the Shannon entropy (or Boltzman-Gibbs entropy which differs from Shannon entropy by a multiplicative constant) with linear constraints.

Instead of specifying a particular entropy measure right from the beginning, we have considered a very general treatment of the entropy maximization in this work and shown that the only entropy measure compatible with the entropy maximization {\`a} la Jaynes is the Shannon entropy if the linear constraints are to be used. In this sense, the procedure devised by Jaynes is strictly devised for the Shannon entropy. As a matter of fact, this has exactly been the point of the well-known Shore-Johnson axioms \cite{Shore}, too. However, we note that we have not used a joint probability composition rule in above derivation thereby rendering our calculations in essence different from the approach of the Shore-Johnson axioms \cite{notelast}.  

When we consider for example the R{\'e}nyi entropy (or Tsallis entropy for that matter) in virtue of Eq. (\ref{eqNew01}), one obtains $0 = (1-q) \beta \frac{\partial \widetilde{U}}{\partial \beta}$. This relation either forces us to use Shannon entropy i.e. setting $q=1$ or assuming $\frac{\partial \widetilde{U}}{\partial \beta} =\frac{\partial U}{\partial \beta}= 0$, which leads to a contradiction since $\frac{\partial p_i}{\partial\beta}\neq0$, as can be seen in Eq. (\ref{eq03a}). Therefore, the use of entropy maximization with linear constraints should not be extended to the uses of the deformed entropies. However, note that our work is limited to the linear constraints i.e. linear averaging schemes so that other averaging schemes is beyond the scope of present treatment.

\begin{acknowledgments}
	T.O. acknowledges the state-targeted program ``Center of Excellence for Fundamental and Applied Physics" (BR05236454) by the Ministry of Education and Science of the Republic of Kazakhstan and ORAU grant entitled ``Casimir light as a probe of vacuum fluctuation simplification" (090118FD5350).
\end{acknowledgments}



\begin{thebibliography}{0}
\bibitem{notelast} The implications of the lack of joint probability composition rule in our derivation deserve closer scrutiny and will be treated elsewhere.
\end{thebibliography}
\end{document}